\documentclass[prl,twocolumn,aps]{revtex4}
\usepackage{longtable,setspace}
\usepackage{graphicx,rotating,lscape}
\usepackage{color}
\usepackage{float,epsfig,amsmath,amssymb,euscript,color}
\usepackage{amsmath,amssymb,epsfig,capt-of,ifthen,calc}
\usepackage{float,euscript}
\usepackage{aecompl}
\usepackage{color}
\usepackage{soul}
\usepackage{multirow}
\usepackage{dcolumn}
\usepackage{longtable}
\usepackage{supertabular}
\usepackage{setspace}
\usepackage{ulem,enumitem}
\usepackage{bm}
\usepackage{hyperref}
\usepackage{natbib}

\usepackage{amsthm}

\definecolor{bleuclair}{rgb}{0.7, 0.7, 1.0}
\definecolor{rosepale}{rgb}{1.0, 0.7, 1.0}

\begin{document}

\title{Large-scale structure and hyperuniformity of amorphous ices}

\author{Fausto Martelli$^{1}$, Salvatore Torquato$^{1,2}$, Nicolas Giovambattista$^{3,4}$, Roberto Car$^{1,2}$}
\affiliation{%
$^{1}$Department of Chemistry, Princeton University, Princeton NJ, USA \\
$^{2}$Department of Physics, Princeton University, Princeton NJ, USA \\
$^{3}$Department of Physics, Brooklyn College of the City University of New York, New York, NY, USA\\
$^{4}$Ph.D. Programs in Chemistry and Physics, The Graduate Center of the City University of New York, New York,
New York 10016, USA
}

\begin{abstract}
  We investigate the large-scale structure of amorphous ices and transitions
  between their different forms by quantifying their large-scale density fluctuations. Specifically, we simulate the
  isothermal compression of low-density amorphous ice (LDA) and hexagonal ice to produce high-density 
  amorphous ice (HDA). Both HDA and LDA are nearly hyperuniform, i.e., 
  they are characterized by an anomalous suppression of large-scale density fluctuations.
  By contrast, 
  in correspondence with the non-equilibrium phase transitions to HDA, the presence 
  of structural heterogeneities strongly suppresses the hyperuniformity and the system 
  becomes hyposurficial (devoid of "surface-area fluctuations"). 
  Our investigation challenges the largely accepted ``frozen-liquid'' picture, which views glasses as structurally 
  arrested liquids. Beyond implications for water, our findings enrich our understanding of pressure induced structural 
  transformations in glasses.
\end{abstract}

\maketitle

\paragraph*{Introduction.--}
Disordered hyperuniform materials are exotic amorphous
states of matter that lie between a crystal and a liquid: they are like
perfect crystals in the way they suppress large-scale density fluctuations
and are like liquids or glasses in that they are statistically
isotropic with no Bragg peaks~\cite{torquato_local_2003}. 
Central to the concept of hyperuniformity is the structure factor $S(\bf{k})$
which, in the thermodynamic limit, is related to $\tilde{h}(\textbf{k})$, the Fourier transform of 
the total correlation function $h(\mathbf{r})$, $S(\mathbf{k})=1+\rho\tilde{h}(\mathbf{k})$.
In $d$-dimensional Euclidean space, $\mathbb{R}^{d}$ ($d=3$ in this work),
the number variance $\sigma^2(R)$ of particles 
inside a spherical window of radius $R$ is related to $S(\mathbf{k})$ via~\cite{torquato_local_2003}: 
\begin{equation}
  \sigma^2(R)=\left<N(R)\right>\left[\frac{1}{(2\pi)^d}\int_{\mathbb{R}^{d}} S(\mathbf{k})\tilde{\alpha}(\mathbf{k};R)d\mathbf{k}\right]
  \label{eq:sigma_1}
\end{equation}
where $\left<N(R)\right>$ is the average number of particles inside the spherical window 
and $\tilde{\alpha}(\mathbf{k};R)$ is the 
Fourier transform of $\alpha(\mathbf{r};R)$, defined as
the volume common to two spherical
windows with centers separated by a vector $\textbf{r}$, divided by the volume of the window.
For a large class of ordered and disordered systems, the number variance has the following large-$R$
asymptotic behavior~\cite{torquato_local_2003,zachary_hyperuniformity_2009}:
\begin{equation}
  \sigma^2(R)=2^d\phi\left[A\left(\frac{R}{D}\right)^d+B\left(\frac{R}{D}\right)^{d-1}+\ell\left(\frac{R}{D}\right)^{d-1}\right] 
  \label{eq:sigma}
\end{equation}
where $\phi=\rho v_1(D/2)$ is the dimensionless density,
$D$ is a characteristic length, $v_1(R)$ is the volume of the spherical window, 
$A$ and $B$ are ``volume'' and ``surface-area'' 
coefficients, respectively, while $\ell\left(R/D\right)^{d-1}$ represents terms of lower order 
than $\left(R/D\right)^{d-1}$. 
$A$ and $B$ can be expressed as:
\begin{equation}
  A=\lim_{|\mathbf{k}|\rightarrow 0}S(\mathbf{k})=1+d2^d\phi\left<x^{d-1}\right>
  \label{eq:A}
\end{equation}
and 
\begin{equation}
  B=-\frac{d^22^{d-1}\Gamma\left(\frac{d}{2}\right)}{\Gamma\left(\frac{d+1}{2}\right)\Gamma\left(\frac{1}{2}\right)}\phi\left<x^d\right>
  \label{eq:B}
\end{equation}
where $x=r/D$, $<x^d>$ is the $d$-th moment of $h(x)$, $<x^d>=\int_{0}^{\infty}x^dh(x)dx$, and $\Gamma(\cdot)$ is the 
gamma function.
In a perfectly hyperuniform system~\cite{torquato_local_2003}, 
the non-negative volume coefficient vanishes, i.e., $A=0$.
Perfect crystals and quasicrystals are exactly hyperuniform.
Disordered hyperuniform systems can be regarded to possess a
``hidden'' long-range order~\cite{torquato_prx}, and have recently been identified in many materials and systems~\cite{donev_unexpected_2005,jiao_avian_2014,pietronero_2002,xie_hyperuniformity_2013,hejna_nearly_2013,florescu_designer_2009,man_isotropic_2013,zito_2015,chremos_2017,zhang_2016,goldfriend_2017}.
On the other hand,
when $A>0$ and $B=0$, the system is \textit{hyposurficial}; examples include 
homogeneous Poisson point
patterns and certain hard-spheres systems~\cite{torquato_local_2003}.
For $A>0$, the ratio $H=S(0)/S(k_{peak})$, where $k_{peak}$ is the wave number
at the largest peak height of $S(\mathbf{k})$,
measures qualitatively how close a system is to perfect hyperuniformity.
Systems in which $H\lesssim10^{-3}$ are deemed to be nearly hyperuniform~\cite{atkinson_2016}. \par
In this Letter, we quantify the large-scale density fluctuations
of amorphous ices modeled via classical molecular dynamics simulations.
The samples contain $N=8192$ water molecules 
described by the classical TIP4P/2005 interaction potential~\cite{abascal_tip4p2005}. 
Following Ref.~\cite{nicolas_pressure_2015}, we consider low-density amorphous ice (LDA) generated by 
quenching the liquid at ambient conditions to a temperature of $80$ K with a rate of $1$ K/ns. 
We then produce two samples of high-density amorphous ice (HDA) by applying pressure with a rate of $0.01$ GPa/ns to 
LDA and to hexagonal ice (I$h$) while keeping the temperature 
constant at T$=80$ K. 
We refer to the HDA produced from I$h$ to as HDA$_{Ih}$, 
and to the HDA produced from LDA to as HDA$_{LDA}$.
We show that all these amorphs are nearly hyperuniform.
In correspondence with the non-equilibrium phase transitions (PTs), structural 
heterogeneities, i.e., molecules in highly distorted local environments, strongly suppress the hyperuniformity and, 
remarkably, the system becomes
hyposurficial. We infer that hyposurficiality is caused by the spatially 
nearly uncorrelated distribution of clusters of such heterogeneities.
We also show that a significant suppression of large-scale density fluctuations emerges when 
cooling water below the temperature 
of freezing of the rotational motions, $T_{rot}$, and we elucidate the connection between molecular rotations 
and large-scale density fluctuations. $T_{rot}$ is lower than the glass transition temperature $T_g$ at which the 
translational motions freeze. \par
To the best of our knowledge, hyposurficiality has never been detected before in any
structural transformation. In the present context, it further supports the notion that the LDA-to-HDA 
transition is first-order-like, consistent with the hypothesized presence of a second critical point in 
this water model~\cite{abascal_2010,singh_2}.
Our findings shed light on 
amorphous ices, which play a pivotal role in understanding water properties, and 
shed also light on structural properties of general glasses. Finally, this work
can stimulate experimentalists to probe $S(\mathbf{k})$ at small 
wave numbers in amorphous systems, and to design experiments to detect large-scale 
structural order.

\paragraph*{Results.--}
At deeply supercooled conditions, water exhibits polyamorphism, i.e., it exists in more
than one amorphous form. The most common forms
are LDA and HDA~\cite{debenedetti_2003,giovanbattista_2006,mishima_1998,loerting_the_2015,aman_colloquium_2016}, 
and a very high density form has also been proposed to exist at even higher pressures~\cite{loerting_second_2001}.
Amorphous ices have been structurally characterized at short- and 
intermediate-length-scale~\cite{tulk_structural_2002,finney_structures_2002,winkel_equilibrated_2011}, 
but no study has probed
their large-scale density fluctuations, even though reported experimental $S(\bf{k})$'s seem to reach 
very small values in the limit of $\mathbf{k}\rightarrow 0$~\cite{tulk_structural_2002,finney_structures_2002}.\par
%
%
\begin{figure}[!]
 \centering
    \includegraphics[scale=.33]{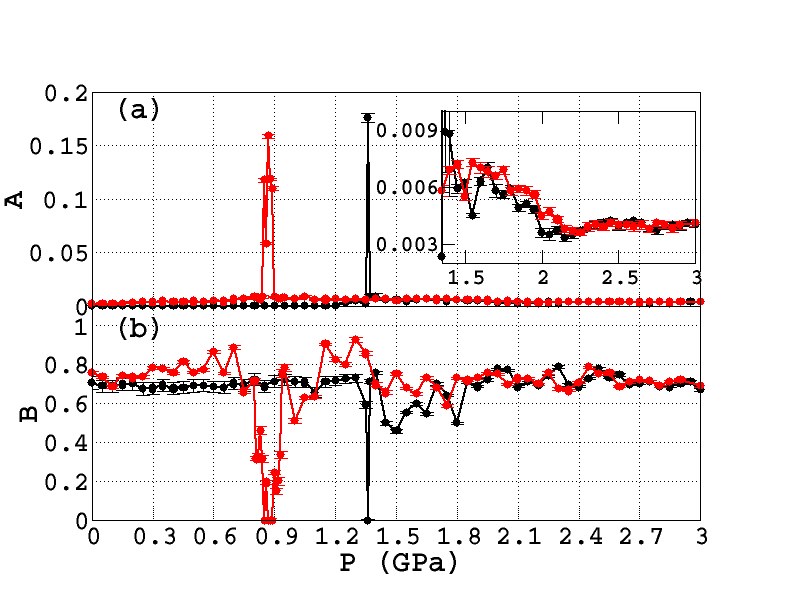}
    \caption{(a) coefficient $A$ during compression of I$h$ (black) and LDA (red). Inset: zoom for $1.4<p<3.0$ GPa. 
             (b) coefficient $B$ during compression of I$h$ (black) and LDA (red). The peaks of $A$ signal the I$h$-to-HDA and 
             LDA-to-HDA transformations, in correspondence of which the system becomes hyposurficial, as shown by the sharp drop of $B$.}
 \label{fig:Fig1}
\end{figure}
Figs.~\ref{fig:Fig1} (a) and ~\ref{fig:Fig1} (b) show, respectively, the coefficients $A$ and $B$ during the 
I$h$-to-HDA$_{Ih}$ and LDA-to-HDA$_{LDA}$ transformations.
The deviation of $A$ 
from zero indicates the degree to which the system departs from hyperuniformity
in light of its connection with $S(\mathbf{k})$ (Eq.~\ref{eq:A}).
Classical crystals at $T=0$ K are trivially hyperuniform and, hence, $A=0$.
Our I$h$ sample acquires slightly larger values, $A\sim10^{-4}$, because of the finite temperature. 
$A$ is 
also low in LDA ($A\sim10^{-3}$). 
In correspondence with the non-equilibrium PTs (at $0.85\lesssim p_T\lesssim1.0$ GPa in LDA, and at 
$1.35\lesssim p_T\lesssim1.36$ GPa in I$h$), $A$ shows a sharp peak, while 
$B$ drops to almost zero ($\lesssim10^{-5}$, Fig.~\ref{fig:Fig1}(b)). 
We infer that hyposurficiality, as indicated by $B\rightarrow 0$, is due to
heterogeneities arranged in clusters distributed in a nearly uncorrelated fashion
(Fig. S1 in Supplemental Material).
The abrupt change in $A$ and $B$
is a static signature of a first-order-like PT.
At high pressures, the predominance of $B$ over $A$ is restored in 
both HDA$_{Ih}$ and HDA$_{LDA}$.
This suggests that the large-scale
density fluctuations are suppressed and are comparable with those in LDA. However, 
the large-scale structures of
HDA$_{Ih}$ and HDA$_{LDA}$ slightly differ one from another at intermediate pressures, 
i.e., at $1.5\lesssim p\lesssim2.2$ GPa, 
but upon further compression, the coefficients A of both HDA's become almost identical (Fig.~\ref{fig:Fig1}, inset), 
suggesting that HDA$_{Ih}$ and HDA$_{LDA}$ have similar large-scale structures.
This observation is further reinforced by the similar values acquired by the coefficient $B$
for both HDA's at $p\gtrsim 2.1$ GPa.\\
\begin{figure}[!]
 \centering
    \includegraphics[scale=.33]{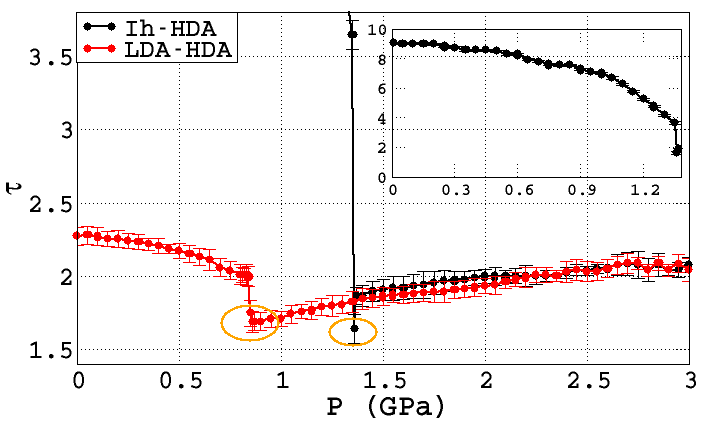}
    \caption{Translational order metric $\tau$ during compression of $I$h (black) and LDA (red). Sudden drops of $\tau$ 
            occur in correspondence of the PTs. Inset: $\tau$ of I$h$ at low pressures.}
 \label{fig:Fig2}
\end{figure}
To get deeper insight into the hyposurficiality and hyperuniformity of amorphous ices, we compute 
the translational order metric $\tau$~\cite{torquato_prx}: 
\begin{equation}
  \tau=\frac{1}{(2\pi)^3\rho^2 D^3}\int_{0}^{\infty}\left[S(\mathbf{k})-1\right]^{2}d\mathbf{k}
  \label{eq:tau}
\end{equation}
where $D$ is a characteristic microscopic length scale.
In the thermodynamic limit, $\tau$ diverges for perfect crystals while it vanishes identically 
for spatially uncorrelated systems. Thus, a deviation of $\tau$ from zero,
which can only be positive, measures the degree of translational order relative to the 
fully uncorrelated case.
In Fig.~\ref{fig:Fig2} we report $\tau$ for the compression of I$h$ (black) and LDA (red). 
Since I$h$ possesses long-range order, large positive values of $\tau$ for I$h$ are removed from the main figure 
and are reported in the inset. 
This representation allows us to: (i) emphasize the lower values of $\tau$ of LDA compared 
to I$h$, and (ii) report the jump in $\tau$ in correspondence with both PTs. 
The finite values of $\tau$ for I$h$ are caused by the finite-size 
of our sample, which causes a truncation of the upper integration limit in Eq.~\ref{eq:tau}
~\footnote{The upper integration limit is 
defined by the reciprocal vectors associated to the natural periodicity of $I_h$ in units of the reciprocal lattice vectors of the 
sample supercell.}. 
For $p<p_T$, $\tau$ decreases continuously upon compression in both samples, as structural heterogeneities appear
which, at this stage, affect $\tau$ but are not concentrated enough to sensitively influence the values 
of $A$ and of $B$.
In correspondence with the PTs,
$\tau$ acquires its minimum values, indicating that some degree of decorrelation should be present. 
\begin{figure}[!]
 \centering
    \includegraphics[scale=.36]{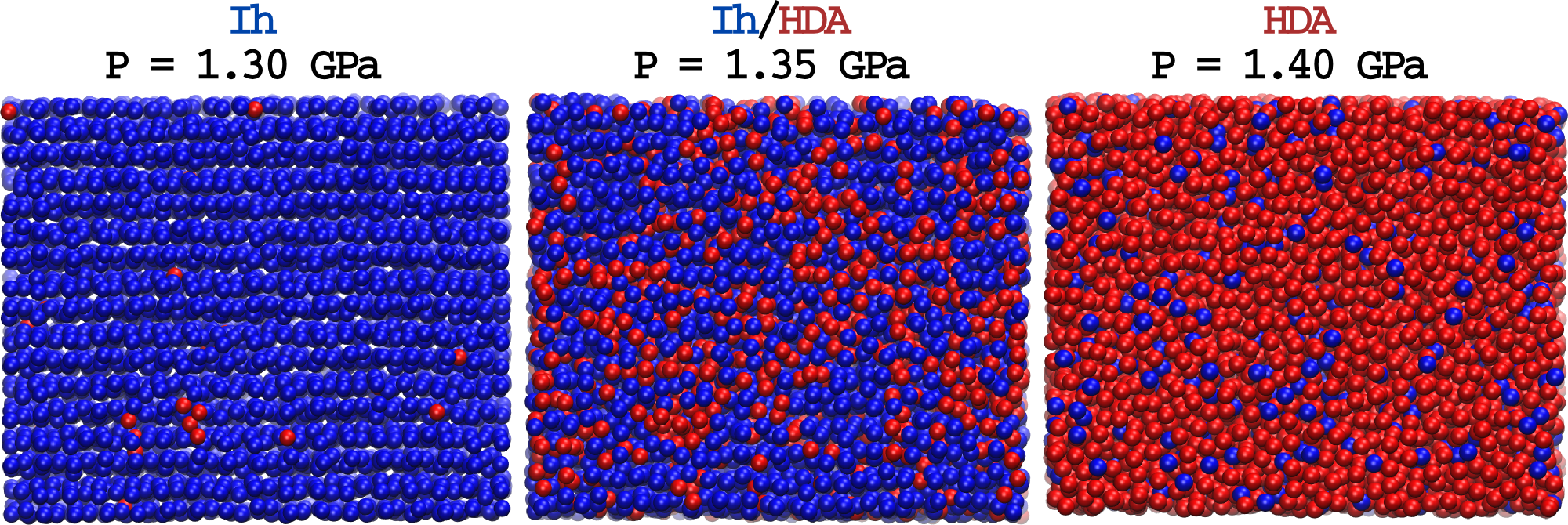}
    \caption{Snapshots of representative configurations before (left), near (center), and after (right) the $I$h-to-HDA transformation. 
             Oxygen sites are represented by blue or red spheres, respectively, for $I$h or HDA character based on the 
             local structure index.}
 \label{fig:Fig3}
\end{figure}
Moreover, $\tau$ shows a discontinuity, further indication of a first-order-like PT.
After the PT, the $\tau$'s of HDA$_{Ih}$ and HDA$_{LDA}$
differ slightly from one another for $1.5\lesssim p\lesssim2.2$ GPa. 
This difference is a further signature of structural differences 
at the large-scale in HDA$_{Ih}$ and in HDA$_{LDA}$. Upon further compression, 
such differences become negligible. 
Notably, the values of $\tau$ for both HDAs at high pressures ($p\sim3.0$ GPa) 
are very close to the values of $\tau$ in LDA for $0.5\lesssim p\lesssim0.8$ GPa. This suggests that
similar translational order is present in both amorphous structures,
in contrast 
with the reported differences at small distances~\cite{soper_structures_2000,finney_structures_2002,nicolas_pressure_2015}. 
At small length scales, the local environment is tetrahedral with well-separated first and second shells of neighbors 
in LDA and I$h$. By contrast, the environment is distorted in HDA, similar to liquid water at standard conditions, due to interstitial 
molecules populating the first intershell region~\cite{soper_structures_2000}.
In Fig.~\ref{fig:Fig3} we report representative snapshots of the I$h$-to-HDA transformation taken 
before ($p=1.30$ GPa), in correspondence with ($p=1.36$ GPa), and after ($p=1.40$ GPa) the 
PT. Blue spheres represent I$h$ sites, red spheres represent HDA environments.
The I$h$-like or HDA-like character is based on 
the local structure index~\cite{shiratani_growth_1996,shiratani_molecular_1998}, which
quantifies the presence of interstitial molecules in the intershell region. 
A similar picture holds for the LDA-to-HDA transformation. \par
\begin{figure}[!]
 \centering
    \includegraphics[scale=.33]{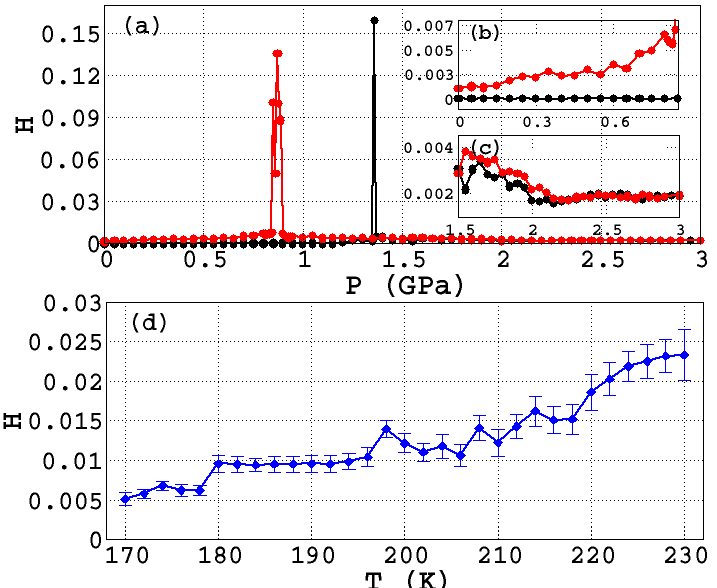}
    \caption{(a) $H=S(0)/S(k_{peak})$ during compression of $I$h (black) and LDA (red). The peaks in correspondence of the 
             PTs (I$h$-to-HDA and LDA-to-HDA) reach values that are two orders of magnitude larger than the qualitative threshold 
             for hyperuniformity ($H\sim10^{-3}$). (b) zoom of $H$ before the LDA-to-HDA PT. (c) zoom of $H$ after the I$h$-to-HDA PT. 
             (d) $H$ during cooling of liquid water at $p=0.1$ GPa (the data are averaged over 10 independent runs).}
 \label{fig:Fig4}
\end{figure}
We quantify the hyperuniformity of the system by calculating the ratio $H=S(0)/S(k_{peak})$. 
Fig.~\ref{fig:Fig4} (a) reports $H$ for the compression of I$h$ (black) and LDA (red).
Fig.~\ref{fig:Fig4} (b) shows that I$h$ has $H\sim10^{-4}$ before the PT, and is, therefore, close to perfect 
hyperuniformity, whereas LDA has $H$ ranging from $\sim2\times10^{-3}$ to $\sim6\times10^{-3}$, signaling that it is nearly hyperuniform. 
The degree of hyperuniformity continuously decreases with increasing pressure, 
due to the appearance of HDA-like sites representing structural heterogeneities.
In correspondence with the PTs, heterogeneities suppress the hyperuniformity, as indicated by the 
spikes of $H$.
At $p>p_T$, HDAs are produced and their values of $H$ are shown in Fig.~\ref{fig:Fig4} (c). Both HDAs 
are nearly hyperuniform, with $H$ ranging from $\sim4\times10^{-3}$ to 
$H\sim2\times10^{-3}$ with increasing pressure. 
The large scale structures of the two amorphous ices, which differ slightly for $p< 2.2$ GPa, become quite similar 
above $2.2$ GPa. At these high pressures the degree of hyperuniformity of the HDAs is similar to that of LDA at low pressure.
Note that LDA and HDA have different hydrogen bond networks (HBNs).
The HBN of LDA is dominated by $5$-, $6$-, and $7$-fold rings, in contrast to HDA, whose HBN 
includes a significant fraction of longer member rings to accommodate the larger density~\cite{martonak}.
Despite these differences, each water molecule is almost perfectly four-fold coordinated in LDA and HDA~\cite{martonak}.
Therefore, the nearly hyperuniform nature 
of LDA and of HDA indicates that the HBNs of both amorphous ices
belong to the class of nearly hyperuniform bond networks~\cite{hejna_nearly_2013}, 
i.e., isotropic networks lacking of crystallinity and coordination defects, 
in which all particles are perfectly coordinated, forming continuous random networks -CRNs-, enriched 
with the suppression of large-scale density fluctuations. \par
The structure factor of an equilibrium liquid in the infinite-wavelength limit
is related to the isothermal compressibility $\kappa_T$ via  
$\lim_{\textbf{k}\rightarrow 0}S(\textbf{k})=\rho k_BT\kappa_T$, where $\rho$ is the density and $k_B$ is the Boltzmann 
constant. The liquid $S(\mathbf{k})$ displays a positive 
curvature for 
small wave-vectors~\cite{Nilsson2014,wikfeldt_spatially_2011,dhabal_probing_2016,clark_structure_2010}.
In our liquid water model at $T=300$ K $H\sim0.07$ and progressively decreases
with cooling down to the supercooled regime (Fig.~\ref{fig:Fig4} (d)). For this 
model, at the adopted cooling rate, 
the glass transition occurs at $T_g\approx200$ K~\cite{nicolas_pressure_2015}. In the temperature 
range $196<T<204$ K, $H\sim0.011$ indicating that, in correspondence with the freezing of the translational degrees of 
freedom, the system is still not hyperuniform. 
Large-scale density fluctuations keep occurring as the sample 
is cooled down to $T_{rot}\sim 180$ K~\cite{martelli_LOM}.
Upon further cooling the large-scale density fluctuations drop rapidly to $H\sim0.006$ immediately below $T_{rot}$ 
and then keep decreasing continuously (data not shown) until they saturate at $H\sim0.002$ for $T\sim110$ K. We attribute 
the large-scale density fluctuations for $T_{rot}\lesssim T\lesssim T_g$ mainly to the changes of the HBN caused by the 
molecular rotations~\cite{martelli_LOM}, while molecular diffusion is the main contributor for $T>T_g$.
The continuous decrease of fluctuations for $T\lesssim T_{rot}$ does not involve any rearrangement of the 
HBN~\cite{martelli_LOM}. It is due instead to small local displacements that reflect the anharmonicity of the potential in the 
glass and conspire to significantly reduce the large-scale density fluctuations when the CRN is fully formed. Thus, the degree 
of hyperuniformity is affected by the amplitude of the vibrational motions. The presence of relaxation processes at temperatures 
well below $T_g$ indicates that LDA is not simply a structurally arrested liquid.\par
The quantity $A/B$ is the ratio of volume to surface-area fluctuations and is a useful metric to detect the 
emergence of hyperuniformity or hyposurficiality at thermodynamic conditions away from criticality; at criticality,  
this ratio becomes meaningless. As shown in Fig. S2 of the Supplemental Material, we find that nearly hyperuniform states 
occur when $A/B<0.015$.
\paragraph*{Conclusions.--}
By analyzing the long wavelength density fluctuations of LDA and HDA generated with classical molecular dynamics 
simulations, we found that both amorphous ices are nearly hyperuniform and have a similar degree of hyperuniformity. 
This suggests that they should possess similar long-range order in spite of their clear differences at the short- and 
intermediate-range scales independently of the preparation protocol followed to produce HDA. In correspondence with the transformation of 
I$h$-to-HDA and of LDA-to-HDA, the applied pressure produces clusters of spatially nearly uncorrelated heterogeneities 
that destroy hyperuniformity. When this occurs, the samples become hyposurficial as density fluctuations that grow 
like the surface-area of the observation window are absent. Hyposurficiality is a static signature that should be inspected 
whenever a first-order PT is involved, which, to our knowledge, was never previously observed as a signature of phase coexistence 
in any context.
The sudden appearance of hyposurficiality, the discontinuous profile of the translational order metric $\tau$, and the spike of $H$ 
in correspondence with the PTs, lead us to conclude that the observed I$h$-to-HDA and LDA-to-HDA transformations are of the 
first kind. The first order nature of the LDA-to-HDA metastable phase transition makes conceivable the existence of a 
second critical point in our model of water.\par 
An additional important finding of our investigation is that the large scale density fluctuations keep decreasing 
well below $T_g$ and $T_{rot}$, i.e., well below the temperature of freeezing of diffusional and rotational motion, 
challenging the notion of glasses as kinetically arrested liquids. Our results also indicate that the degree of hyperuniformity 
of a glass is affected by vibrational motion and, in particular, that not all glasses are hyperuniform. \par
Finally, we propose that away from criticality, the ratio $A/B$ could provide a useful metric to gauge the degree of volume 
to surface-area fluctuations, which include hyperuniform and hyposurficial systems at the extremes.\par

\begin{acknowledgments}
F. M. and R. C. acknowledge the Department of Energy (DOE), grant number DESC0008626
\end{acknowledgments}

\bibliography{References}

\end{document}


\title{Supplementary Material for: Large-scale structure and hyperuniformity of amorphous ices}

\author{Fausto Martelli$^{1}$, Salvatore Torquato$^{1,2}$, Nicolas Giovambattista$^{3,4}$, Roberto Car$^{1,2}$}
\affiliation{%
$^{1}$Department of Chemistry, Princeton University, Princeton NJ, USA \\
$^{2}$Department of Physics, Princeton University, Princeton NJ, USA \\
$^{3}$Departament Physics, Brooklyn College of the City University of New York, New York, NY, USA\\
$^{4}$Ph.D. Programs in Chemistry and Physics, The Graduate Center of the City University of New York, New York,
New York 10016, USA
}

\maketitle

\begin{figure}[!]
 \centering
    \includegraphics[scale=.33]{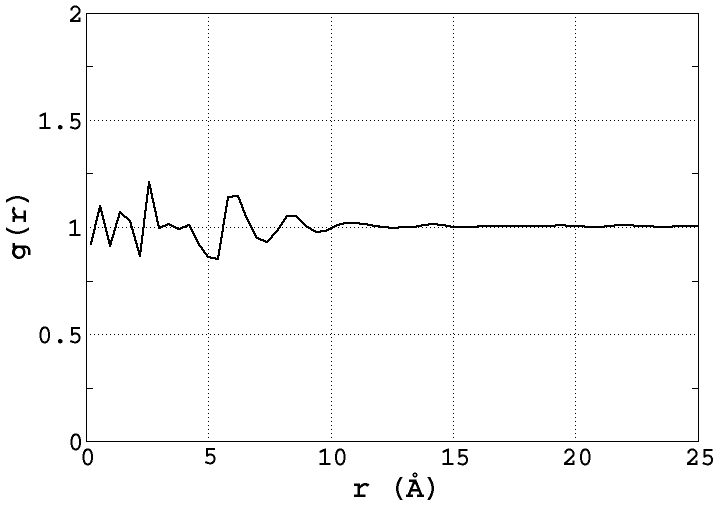}
    \caption{Radial distribution function of the centers of mass of the heterogeneous clusters. Here we consider clusters of 
     red spheres (see center panel of Fig.3). A cluster contains at least two neighboring sites of same character. A similar
     figure could have been drawn using the blue spheres in Fig.3.}
 \label{fig:SM1}
\end{figure}
 \begin{figure}[H]
 \centering
    \includegraphics[scale=.33]{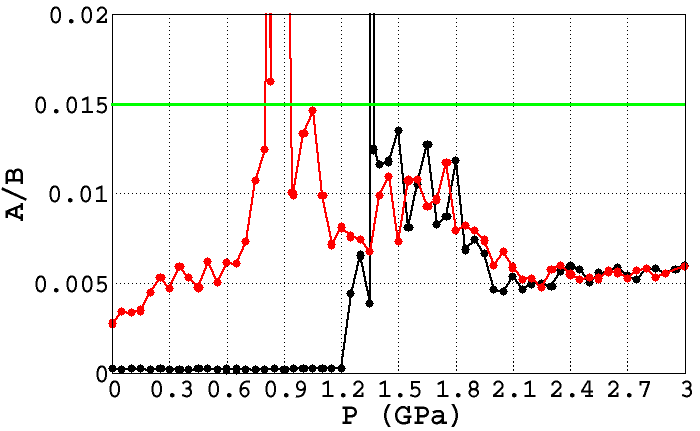}
    \caption{Ratio $A/B$ during compression of I$h$ (black) and LDA (red). In correspondence with the structural transformations of I$h$ 
             and LDA to HDA, the ratio $A/B$ peaks at very high values of $\sim10^4$, i.e. well outside the range of the figure where 
             only values of $A/B$ up to $0.02$ are reported. The horizontal green line indicates a qualitative threshold below 
             which a system can be considered to be approximately hyperuniform. The value of $A/B$ fluctuates near $\sim0.0002$ for I$h$, 
             while it fluctuates near $\sim0.005$ for LDA and HDA away from the regions containing a larger fraction of heterogeneities 
             below and above the peaks associated to the structural transformations. We speculate that it should be possible to 
             prepare amorphs with higher degree of hyperuniformity with e.g. protocols having slower cooling/pressurizing rates.}
 \label{fig:SM2}
\end{figure}